\documentclass[aip,apl,amsmath,amssymb,reprint]{revtex4-1}

\usepackage{graphicx}
\usepackage{dcolumn}
\usepackage{bm}

\usepackage{hyperref}
\usepackage{color}
\usepackage{times}

\newcommand{\T}{\mathcal{T}}
\newcommand{\M}{\mathbf{M}}
\renewcommand{\P}{\mathcal{P}}
\newcommand{\E}{\mathcal{E}} 
\newcommand{\A}{\mathcal{A}}
\newcommand{\D}{\mathcal{D}}
\newcommand{\X}{\mathcal{X}}
\newcommand{\V}{\mathcal{V}}
\newcommand{\R}{\mathcal{R}}
\renewcommand{\aa}{\alpha}
\newcommand{\bb}{\beta}
\renewcommand{\gg}{\gamma}
\newcommand{\dd}{\delta}
\newcommand{\oo}{\omega}
\newcommand{\ii}{i}

\allowbreak
 
\begin{document}

\title{Emitter and absorber assembly for multiple self-dual operation and
directional transparency}

\author{P.~A.~Kalozoumis}
 \email[]{pkalozoum@phys.uoa.gr}
\affiliation{Department of Physics, University of Athens, 15771 Athens, Greece}

\author{C.~V.~Morfonios}
\affiliation{Zentrum f\"ur Optische Quantentechnologien, Universit\"{a}t Hamburg, 22761 Hamburg, Germany}

\author{G.~Kodaxis}
\affiliation{Department of Physics, University of Athens, 15771 Athens, Greece}

\author{F.~K.~Diakonos}
\affiliation{Department of Physics, University of Athens, 15771 Athens, Greece}

\author{P.~Schmelcher}
\affiliation{Zentrum f\"ur Optische Quantentechnologien, Universit\"{a}t Hamburg, 22761 Hamburg, Germany}
\affiliation{The Hamburg Centre for Ultrafast Imaging, Universit\"{a}t Hamburg, 22761 Hamburg, Germany}

\date{\today}

\begin{abstract}
We demonstrate how to systematically design wave scattering systems with simultaneous coherent perfect absorbing and lasing operation at multiple and prescribed frequencies. The approach is based on the recursive assembly of non-Hermitian emitter and absorber units into self-dual emitter-absorber trimers at different composition levels, exploiting the simple structure of the corresponding transfer matrices. 
In particular, lifting the restriction to parity-time-symmetric setups enables the realization of emitter and absorber action at distinct frequencies and provides flexibility with respect to the choice of realistic parameters.
We further show how the same assembled scatterers can be rearranged to produce unidirectional and bidirectional transparency at the selected frequencies.
With the design procedure being generically applicable to wave scattering in single-channel settings, we demonstrate it with concrete examples of photonic multilayer setups.
\end{abstract}


\maketitle

\textit{Introduction.}---Control of wave amplification and attenuation is  crucial for a multitude of contemporary technological applications ranging from sensors~\cite{Fleury2015}, filters~\cite{Agarwal2016} and acoustic absorbers~\cite{Merkel2015} to information processing~\cite{Lvovsky2009}. In particular, \textit{lasing} is a landmark phenomenon signified by the induced coherent wave emission when a gain threshold value is reached. First established for optics~\cite{Siegman1989}, lasing can be regarded as a general effect related to active wave media. Wave absorption, on the other hand, plays a significant role in a multitude of applications, such as optical fibers~\cite{Warken2007OE}, solar cells~\cite{Nakayama2008APL,Pala2008AdvMater} and acoustic metamaterials~\cite{Jimenez2016APL}. In order to effectively manipulate absorption, the control of losses is essential. Towards this aim, several approaches have been reported~\cite{Kats2013NatMater,Regensburger2012Nat}, often requiring media with large attenuation which may result in practical limitations~\cite{Villinger2015OL}. In this context, of particular importance is the phenomenon of \textit{coherent perfect absorption} (CPA)~\cite{Chong2010,Jin2016SciRep,Smaali2016SciRep} which allows for very high absorption in structures comprized of materials with low intrinsic losses. A structure operating as a CPA at a certain frequency yields vanishing reflection and transmission for particular modes with waves incident on both sides, while  unidirectional incidence leads to partial absorption. A simple and elegant explanation of CPA was provided in Ref.~\cite{Chong2010}, where it was interpeted as the time-reversed counterpart of a laser.

Gain and loss of wave amplitude at a given frequency can be modeled by a complex `potential' entering the underlying Helmholtz equation \cite{Makris2008PRL}, and expressed by the non-unitarity of a system's scattering ($S$-) matrix. 
In this sense, non-Hermitian scattering is characterized by the magnitude of the  $S$-matrix eigenvalues $s_\pm$ \cite{Ambichl2013} defined below. For the two aforementioned cases, namely lasing and CPA, one of the eigenvalues $s_\pm$ is diverging and vanishing, respectively. Moreover, non-Hermitian scattering has been related to intriguing transmission features such as \textit{unidirectional} or \textit{bidirectional} transparency \cite{Lin2011,Mostafazadeh2013}.

The observation that lasing and CPA can occur at the same frequency $\oo$ for a single setup \cite{Longhi2010} has triggered intensive research activity on possible realization of the CPA-laser condition \cite{Mostafazadeh2009,Longhi2010pra} $|s_-(\oo)| = 0 = 1/|s_+(\oo)|$.
This \textit{self-dual spectral singularity}~\cite{Mostafazadeh2012} is usually associated with the invariance of a system under the combined action of spatial reflection $\P$ and time reversal $\T$~\cite{Longhi2010,Stone2011,Schindler2012,Ambichl2013}. The CPA-laser condition indeed corresponds to vanishing of both diagonal elements $\alpha$ and $\delta$ of the system’s transfer matrix (TM) (Eq.\ (\ref{tm1})),which are related by complex conjugation under $\P\T$ symmetry~\cite{Longhi2010}, meaning that CPA and laser modes always occur at the same frequency in $\P\T$-symmetric setups. Nevertheless, they lie in the $\P\T$ broken phase~\cite{Stone2011}, indicating that $\P\T$ symmetry is not essential for self-dual singularities, as shown recently in \cite{Mostafazadeh2012} for a dimer model.
Frequency separation of CPA and laser in non-$\P\T$-symmetric setups may actually be of advantage for the experimental observation of a CPA mode:
slightest deviation from it will involve portion of the dual laser mode which dominates due to diverging $|s_+(\oo)|$.
Thus, accessing the CPA in $\P\T$-symmetric systems would demand extreme precision of setup parameters and input wave amplitudes.

\begin{figure}[t!]
\centering
\includegraphics[width=1\columnwidth]{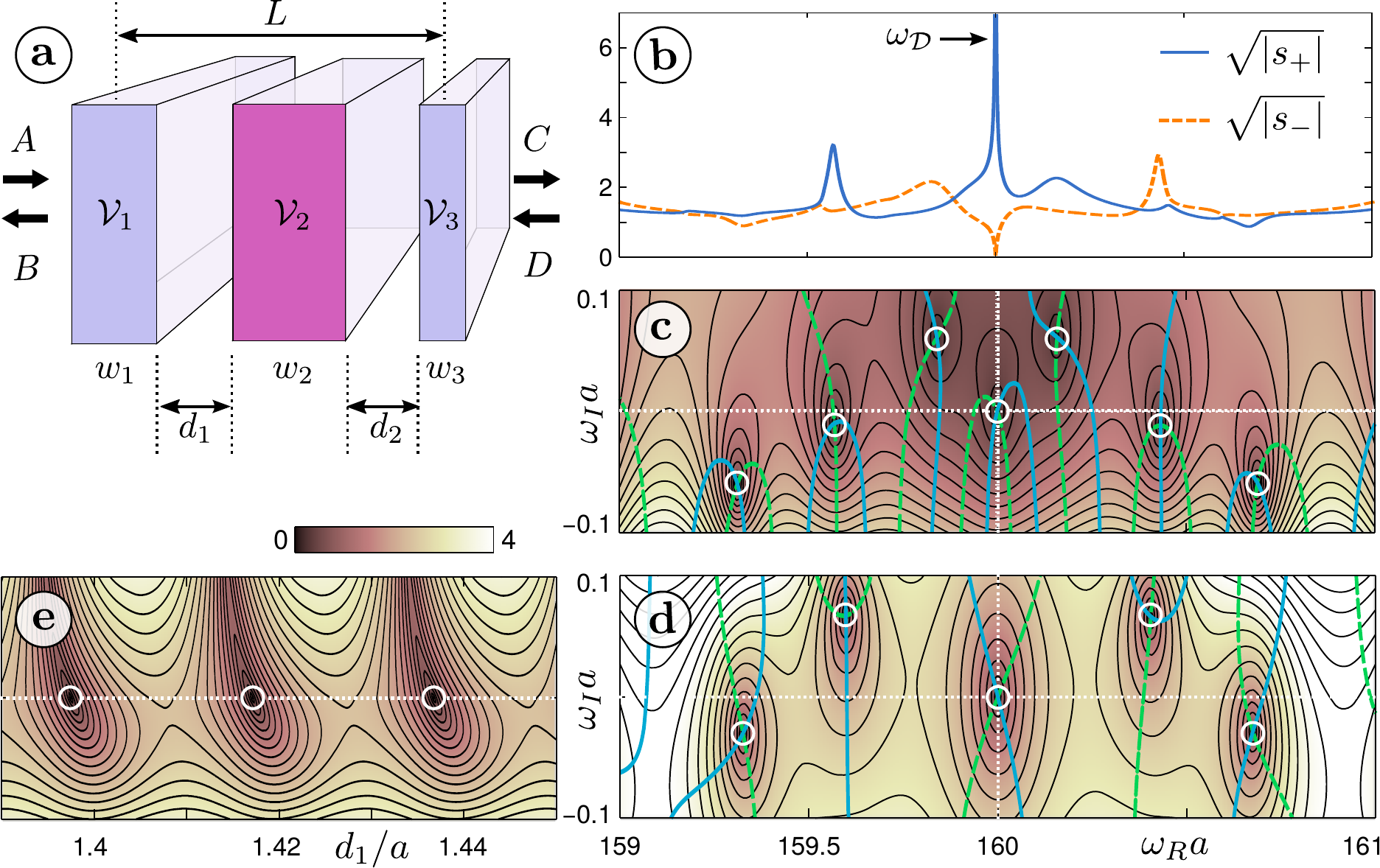}
\caption{ \label{fig1}  (Color online) 
\textbf{(a)} Schematic of a general non-Hermitian trimer, implemented as three optically active dielectric slabs $\V_j$ ($j = 1,2,3$) with refractive indices $n_j$, gain/loss parameters $g_j$, and widths $w_j$, in ambient vacuum (with $n_{0}=1$).
The system acts as self-dual emitter-absorber mode at the selected real frequency $\oo_{\D} = 160 / a$ for  
$w_1 = w_3 = a$, $n_1 = n_3=2.395,~g_1 = g_3 = -0.0056$, and $w_{2} = 2a$, $n_2 = 2.69,~g_{2}=0.0024$, with inter-slab distances $d_{1} = 1.3975 a$ and $d_{2} = 1.3710 a$.
\textbf{(b)}  Magnitude of $S$-matrix eigenvalues for varying real frequency $\oo = \oo_R$ in the vicinity of $\oo_{\D}$. 
\textbf{(c,\,d)} Contour plot of absolute transfer matrix elements (c) $|\alpha(\oo)|$ and (d) $|\delta(\oo)|$ in the complex $\oo$-plane, with zeros (white circles) indicated by intersections of nodal lines of $\alpha_R,\delta_R$ (solid blue) with those of $\alpha_I,\delta_I$ (dashed green), respectively. 
\textbf{(e)}  Contour plot of $|\alpha| + |\delta|$ for varying $\oo_{I}$ and $d_{1}$ at $\oo_R = \oo_{\D}$, for fixed $d_{2}$ (zeros indicated by circles).
} 
\end{figure}

For a one-dimensional setup with $N$ subparts $\V_j$ ($j = 1,2,\dots,N$), denoted by $\V = \V_1\V_2\cdots\V_N$, stationary scattering is described by
\begin{equation} \label{tm1} 
\begin{pmatrix} A  \\ B  \end{pmatrix}
= \prod_{j=1}^{N} \M_j 
\begin{pmatrix} C  \\ D  \end{pmatrix},
~~\M_j =  \begin{pmatrix} \aa_j & \bb_j  \\ \gg_j & \dd_j \end{pmatrix},
\end{equation}
where $A (B)$ and $D (C)$ are the ingoing (outgoing) plane wave amplitudes on the left and right, respectively (see Fig.\,\ref{fig1}), and $\M_j$ is the TM of the $j$-th unit.
In terms of the total transfer matrix $\M = \prod_{j=1}^{N} \M_j \equiv \footnotesize \arraycolsep=1pt\def\arraystretch{.6} \begin{pmatrix} \aa & \bb  \\ \gg & \dd \end{pmatrix}$, the condition for the setup to act as an emitter $\E$ or an absorber $\A$ at a frequency $\oo = \oo_R + i\oo_I$ is $\aa(\oo)=0$ or $\dd(\oo)=0$, respectively \cite{Longhi2010}.
The mapping to a pole $1/|s_+(\oo)| = 0$ or zero $|s_-(\oo)| = 0$ of the $S$-matrix $\mathbf{S} = \frac{1}{\aa} \footnotesize \arraycolsep=1pt\def\arraystretch{.6} \begin{pmatrix} \gg & 1  \\ 1 & -\beta \end{pmatrix}$ is here provided by its determinant $s_+ s_- = - \dd/\aa$, using that det$(\M)=1$. 
Self-dual action requires $\aa(\oo) =  \dd(\oo) = 0$.
From here on we write $\V_j \sim \X [\oo]$, with $\X = \E,\A$ or $\D$, for a unit $\V_j$ which acts as an emitter ($\aa_j=0$), absorber ($\dd_j=0$), or self-dual emitter-absorber ($\aa_j=\dd_j=0$) at $\oo$, respectively. Note that the $S$-matrix eigenvalues can be expressed in terms of the TM elements as
$ s_{\pm}=\frac{1}{2\alpha}\left(-\beta+\gamma \pm \sqrt{4+(\beta+\gamma)^2}\right)$
(we use $c=1$ for the speed of light throughout  the text).

Within a generic wave mechanical framework, let us define a scatterer to act as an `emitter' when $\alpha(\oo) = 0$ at a certain frequency $\oo$, as `absorber' when $\delta(\oo) = 0$, and as a `self-dual' emitter-absorber when $\alpha(\oo) = \delta(\oo) = 0$.
This corresponds to an optical setup supporting a laser mode, a CPA mode, or both, respectively.
The proof of principle in Ref.~\cite{Mostafazadeh2012} suggests that non-$\P\T$-symmetric scatterers could be more generally employed for the realization of emitter and absorber action at \textit{common} or at \textit{separate} frequencies, a scenario investigated very recently~\cite{Xiao2016NJP, Hang2016NJP}. 
Even when surpassing the restrictions imposed by $\P\T$ symmetry, however, it quickly becomes challenging to design emitter, absorber, or, in particular, self-dual setups at \textit{multiple prescribed} frequencies, which at the same time possess realizable geometric and material characteristics.
Indeed,  without consulting a natural design principle, the direct fulfillment of the above TM conditions for a multiparametric system at real scattering frequencies---by locating isolated solutions compatible to the physical constraints---would demand a high algorithmic complexity and computational cost.

In this Letter we propose a systematic procedure for the assembly of self-dual scattering setups at multiple prescribed frequencies from non-Hermitian subparts, based on recursive combination of units, easily set to act as emitters or absorbers at the desired real frequencies with realistic setup characteristics.
This technically separates the determination of needed unit parameters from their spatial configuration, drastically simplifying the total system design, with added flexibility by lifting the constraint of $\P\T$ symmetry.
We further show how selective unidirectional and bidirectional transparency is enabled by rearranging the same emitter, absorber, or self-dual scattering units.
The method is illustrated for photonic multilayer setups.

{\it Design of a trimer.}---The aim is to develop a simple yet systematic concept to design a setup exhibiting self-dual action at a preselected \textit{real} frequency.
For the simplest composite setup with $N=2$ units, we have $\M=\M_1\M_2$ with $\aa = \aa_1\aa_2 + \bb_1\gg_2$ and $\dd = \dd_1\dd_2 + \gg_1\bb_2$.
Notice that a self-dual setup $\V \sim \D[\oo]$ \textit{cannot} be assembled by the combination of individual emitter and absorber units $\V_j \sim \E[\oo]$ or $\V_j \sim \A[\oo]$ ($j=1,2$), since this would require $\bb_1\gg_2 = \gg_1\bb_2 = 0$, making $\M_1$ and/or $\M_2$ non-invertible ($\det \M_j = 0$).
Thus, to achieve a two-unit self-dual setup, one must find solutions of $\aa = \dd = 0$ by combining the internal characteristics of the units $\V_1$ and $\V_2$ with the frequency variable $\oo$.
Although this may be possible in principle, it relies on fine tuning the available parameters subject to constraints.
For example, if the units are optically active dielectric slabs, solutions are sought for realizable slab widths and spacing, refractive indices, as well as loss and gain rates, while keeping $\oo$ on the real axis in the complex plane.
Such constraints may be very challenging to meet by directly imposing the self-dual condition on the total TM due its complex dependence on the parameters.

Let us now demonstrate how self-dual setups are obtained systematically for any preselected real frequency $\oo$ from suitable \textit{combinations of emitters and absorbers}. 
The principle is based on considering three units $\V_j$ ($j=1,2,3$) assembled into a setup $\V = \V_1\V_2\V_3$, as shown in of Fig.\,\ref{fig1}\,(a), whose total transfer matrix $\M=\M_1\M_2\M_3$ has diagonal elements
\begin{subequations} 
\begin{align} 
\label{eq:TM11_3unit} 
\aa &= \aa_1 (\aa_2\aa_3 + \bb_2\gg_3) + \bb_1 (\gg_2\aa_3 + \dd_2\gg_3), \\
\label{eq:TM22_3unit} 
\dd &= \gg_1 (\aa_2\bb_3 + \bb_2\dd_3) + \dd_1 (\gg_2\bb_3 + \dd_2\dd_3).
\end{align}
\end{subequations}
In contrast to the dimer considered above, this trimer setup can be rendered self-dual, $\V \sim \D[\oo]$, by \textit{separately} making the units $\V_j$ emitters or absorbers at $\oo$, and then determining their required spacing.
Indeed, the emitter condition $\aa(\oo) = 0$ is fulfilled by imposing $\aa_1=\aa_3 =\dd_2 = 0$ in Eq.\,(\ref{eq:TM11_3unit}), that is, by rendering $\V_{1,3} \sim \E[\oo]$ and $\V_2 \sim \A[\oo]$.
Note that, at the level of the separate transfer matrices $\M_j$, which have a simple parametric dependence, this task is easily achieved for realistic parameters.
Simultaneous absorber action at $\oo$ additionally requires $\dd(\oo)=0$, yielding  
\begin{equation}
\label{eq:absorber_condition} 
 \aa_2 = -\frac{\gg_2}{\gg_1}\dd_1 -\frac{\bb_2}{\bb_3}\dd_3  = -\left(\frac{\gg_2}{\gg_1} + \frac{\bb_2}{\kappa\bb_1} \right)\dd_1  \\
\end{equation}
where, for simplicity, in the second equality the unit $\V_3$ has been set identical to the unit $\V_1$ while shifted by $L$ (see Fig.\,\ref{fig1}\,(a)), with $\kappa = e^{-2ikL}$ being the corresponding phase shift for the vacuum wave vector $k$.
Note that the spatial configuration of the scatterers $\V_j$ is implicitly present in $\M$ through similar phase shifts (from the common origin) entering the $\M_j$.
With given parameters for the separate $\V_j$, the complex Eq.\,(\ref{eq:absorber_condition}) determines their two relative positions so as to make the emitter also an absorber at the same selected $\oo$.

For definiteness, let us apply the above principle to non-Hermitian photonic multilayer structures (at normal light incidence) which are simply parametrized:
Each unit $\V_j$ in the trimer is a homogeneous `slab' of width $w_j$ with complex `refractive index' $n_j + i g_j$ ($g_j$ being the loss/gain rate), at distance $d_j$ from the next slab $\V_{j+1}$, as depicted in Fig.\,\ref{fig1}\,(a).
We here consider the structure $\V = \V_1\V_2\V_3 = \V_1\V_2\V_1$ with the third slab shifted by $L = w_1 + w_2 + d_1 + d_2$ with respect to the first.
Assuming that the slab parameters have been adjusted such that $\V_{1,3} \sim \E[\oo]$ and $\V_2 \sim \A[\oo]$ (by fulfilling $\aa_1=\aa_3 =\dd_2 = 0$), to render the trimer self-dual, $\V \sim \D[\oo]$, we impose the condition (\ref{eq:absorber_condition}) and solve for $d_1 \pm d_2$.
This yields
\begin{subequations} \label{eq:d1_pm_d2}
\begin{align} 
d_1 + d_2 =& \frac{1}{k}\left[ \arctan\left( \frac{\textrm{Im} \lambda}{\textrm{Re} \lambda} \right)+ m\pi \right] \equiv \frac{1}{k} \, f_m(\lambda), 
\label{eq:d1_m_d2} \\
d_1 - d_2 =& \frac{1}{k}\left[\pm \arccos\left( \pm \frac{\textrm{Re}\lambda}{2 \cos f_m(\lambda)} \right) +2m'\pi \right] 
\label{eq:d1_p_d2}
\end{align}
\end{subequations}
with $m,m' \in \mathbb{Z}$ and signs $\pm$ chosen independently, where the parameter $\lambda \equiv (\aa_2/\dd_1) e^{-ik(w_1+w_2)}$ is independent of $d_{1},~d_{2}$ ($\aa_j,\dd_j$ are unaffected by slab shifts).
It was here used that each slab itself is reflection symmetric.
Solving Eq.\,(\ref{eq:absorber_condition}) for $|\lambda|$ further leads to the compact relation $\left| \aa_2/\dd_1 \right| = |\lambda| = 2 \, | \cos k(d_1-d_2) |$ between the remaining (nonzero) diagonal TM elements of the $\V_j$ and their spacing.
This reveals how lifting the restriction of mirror symmetry for the trimer facilitates the design of a self-dual setup:
For symmetric geometry, $d_1 = d_2$, the condition $|\aa_2| = 2 |\dd_1|$ imposes a stringent relation between the slab parameters (while already having demanded $\aa_1 = \dd_2 = 0$) realizable.
On the contrary, when $d_1 \neq d_2$ the slabs can be designed individually as emitter/absorber units at the preselected real $\oo$, with the $d_1 \pm d_2$ obtained from Eqs.\,(\ref{eq:d1_pm_d2}) in combination with a $|\lambda|$ in a whole available range $0 < |\lambda| \leqslant 2$. The  crucial role of the distance between adjacent parts of a structure on the realization of spectral singularities has been recently reported for $\mathcal{PT}$ symmetric systems in Ref.~\cite{Mostafazadeh2016AoP}. Note also that the selection of the ambient medium with refractive index $n_{0}=1$ is not restrictive. The same procedure could have been followed for different embedding media.

In Fig.\,\ref{fig1}\,(b) the $S$-matrix eigenvalues of such a trimer setup are shown, varying $\oo = \oo_R$ across a self-dual point $\oo_\D$ where a pole $|s_+| \to \infty$ and a zero $|s_-| = 0$ coalesce.
The slab widths $w_1 = w_2/2 = a$ are also set as preselected parameters ($a$ being our length unit) along with $\oo_\D$.
The condition $\V \sim \D[\oo_\D \in \mathbb{R}]$ is confirmed in Figs.\,\ref{fig1}\,(c) and (d) where zeros of both $\aa$ and $\dd$ are seen, respectively, to occur on the real axis at $(\oo_R,\oo_I) = (\oo_\D,0)$ in the complex $\oo$-plane.
Several $S$-matrix poles ($\aa = 0$) and zeros ($\dd = 0$) are also seen to occur separately at complex $\oo$-values.
Note here that, in contrast to $\P\T$-symmetric setups \cite{Stone2011}, those do not occur in complex conjugate pairs.

The pole and zero structure of the $S$-matrix in the complex plane is generally modulated by varying one or more parameters of the setup, giving rise to possible coalescence points---and thereby self-dual action---at the same frequency for an alternative setup.
Following the above procedure for the trimer design, this recurrence of self-dual singularities is made controllable by a single spacing parameter.
Indeed, adding or subtracting Eqs.\,(\ref{eq:d1_m_d2}) and (\ref{eq:d1_p_d2}), we see that the self-dual condition $\aa = \dd = 0$ is fulfilled for different values of $d_1$ or $d_2$, respectively, corresponding to the set of integers $m,m'$.
This is illustrated in Fig.~\ref{fig1}\,(e), where vanishing of the quantity $|\aa(\oo)| + |\dd(\oo)|$ is used as an indicator for self-dual action:
As the slab distance $d_1$ is varied, the setup is rendered an emitter-absorber $\V \sim \D[\oo_\D]$ periodically at intervals $\Delta d_1 = \pi/k = 0.02\,a$, with the zeros shown (from the left) corresponding to pairs $(m,m') = (141,1)$, $(140,1)$, $(141,2)$.
The method thus not only enables the design of $\D$ modes at selected $\oo$, but also provides further flexibility with the slab spacings ($d_1$ or $d_2$) as additional degrees of freedom in order to e.\,g., overcome fabrication restrictions.

\begin{figure}
\centering
\includegraphics[width=1\columnwidth]{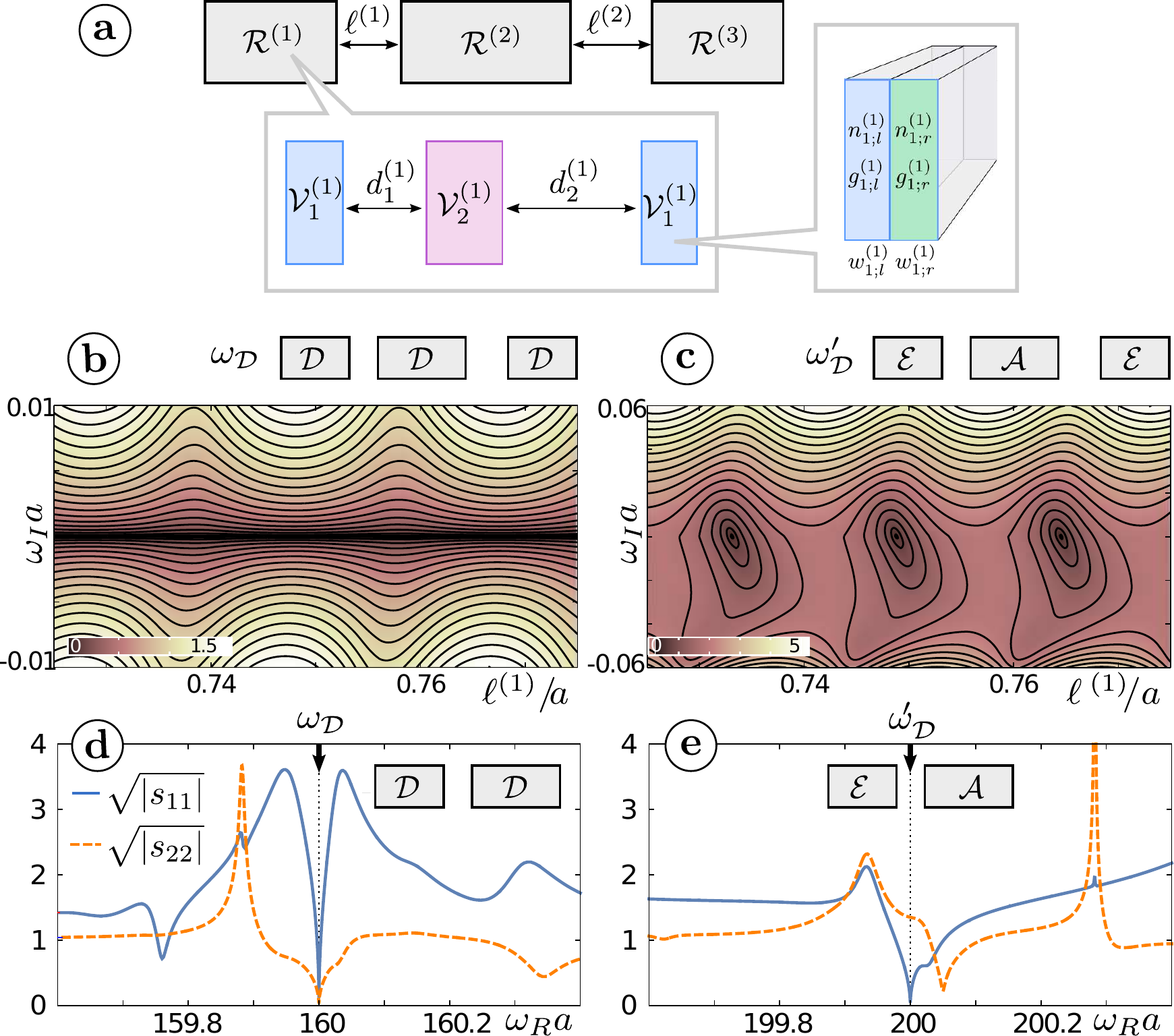}
\caption{\label{fig2}  (Color online) 
\textbf{(a)} Non-Hermitian supertrimer scattering setup $\R = \R^{(1)}\R^{(2)}\R^{(3)}$ implemented as a photonic multilayer and designed to operate as a self-dual emitter absorber at two frequencies $\oo_\D = 160/a$ and $\oo_\D' = 200/a$ for a chosen length unit $a$.
Each superunit $\R^{(\ii)}$ ($i=1,2,3$) is a trimer $\V^{(\ii)}_1\V^{(\ii)}_2\V^{(\ii)}_3$ of bilayers $\V^{(\ii)}_j$ with material parameters $n^{(\ii)}_{j;l[r]},g^{(\ii)}_{j;l[r]}$ for the left [right] layer.
The chosen setup has preselected slab widths $w^{(\ii)}_{j;l} = w^{(\ii)}_{j;r}$ (common among the $\R^{(\ii)}$) and $\R^{(1)} = \R^{(3)}$, $\V^{(\ii)}_1 = \V^{(\ii)}_3$.
\textbf{(b)} Contour plots of $|\alpha| + |\delta|$ with varying $\oo_{I}$ and $\ell^{(1)}$ for $\oo_R = \oo_{\D}$, where $\R^{(1,2,3)} \sim \D[\omega_\D]$, thus making $\R$ self-dual for any $\ell^{(1)}$ at $\oo_{I} = 0$.
\textbf{(c)} Same as (b) but for $\oo_R = \oo_\D'$, where $\R^{(1,3)} \sim \E[\omega_\D']$, $\R^{(2)} \sim \A[\omega_\D']$, making $\R$ self-dual recurrently in $\ell^{(1)}$.
\textbf{(d,\,e)} Reflection coefficients with varying real frequency $\oo = \oo_R$ for the superdimer $\R^{(1)}\R^{(2)}$, showing (d) bidirectional transparency at $\oo_\D$ and (e) unidirectional transparency $\oo_\D'$. 
With slab widths selected as $w^{(1)}_{1;l} = 0.8\,w^{(1)}_{2;l} = a$, the remaining setup parameter values are determined as described in the text and they are:  $d_{1,2}^{(1)} = [0.71,0.69]\,a$,~ 
$d_{1,2}^{(2)} = [0.687,0.685]\,a$,
$\ell^{(1,2)} = [0.75,0.77]\,a,$ 
$[n_{1;l}^{(1)},n_{1;r}^{(1)},n_{2;l}^{(1)},n_{2;r}^{(1)}]= [3.47, 2.50, 1.98, 2.50],$ 
$[n_{1;l}^{(2)},n_{1;r}^{(2)},n_{2;l}^{(2)},n_{2;r}^{(2)}]=[3.19, 1.70,1.61,2.38],$ 
$[g_{1;l}^{(1)},g_{1;r}^{(1)},g_{2;l}^{(1)},g_{2;r}^{(1)}]= [0.78,-4.80,7.43,-1.96]\times 10^{-3}$,
$[g_{1;l}^{(2)},g_{1;r}^{(2)},g_{2;l}^{(2)},g_{2;r}^{(2)}] = [4.53,7.11,7.89,-5.08]\times 10^{-3}$.
}
\end{figure}

{\it Self-dual modes at multiple selected frequencies.}---The design of setups with self-dual operation at \textit{multiple} frequencies requires a higher dimensional parameter space, and the direct search of relevant solutions, without a physical guiding principle, ultimately faces a prohibitive complexity.
We approach this challenge by applying the above concept of combined emitter and absorber units \textit{recursively} at the level of composite scatterers.
The procedure is demonstrated for the assembly of a composite setup with self-dual operation at two preselected real frequencies $\oo_\D$ and $\oo_\D'$.
It consists of three superunits $\R^{(\ii)}$ ($\ii=1,2,3$) at spacings $\ell^{(1)},\ell^{(2)}$, each of which is an extended trimer of units $\V_j^{(\ii)}$ ($j=1,2,3$) at spacings $d_1^{(\ii)},d_2^{(\ii)}$, (Fig.~\ref{fig2}\,(a)).
Like in the simple trimer case above, $\R^{(1)} = \R^{(3)}$ and $\V_1^{(\ii)} = \V_3^{(\ii)}$. 

Applying the same assembly concept at the level of the supertrimer $\R = \R^{(1)}\R^{(2)}\R^{(1)}$, there are two combinations which render it self-dual at the two different $\oo$:
(i) Either each superunit is made self-dual, $\R^{(\ii)} \sim \D[\omega_\D]$ ($\ii = 1,2,3$), so that also $\R \sim \D[\omega_\D]$ [recalling that the product of an even (odd) number of anti-diagonal matrices $\M^{(\ii)}$ is (anti-) diagonal], or
(ii) the central superunit is made absorber, $\R^{(2)} \sim \A[\omega_\D']$, and the peripheral ones emitters, $\R^{(1,3)} \sim \E[\omega_\D']$ (or vice versa), as the units of the simple trimer above, though at the level of superunits.
Considering dielectric slabs, in case (i) the spacings $d_1^{(\ii)},d_2^{(\ii)}$ in each $\R^{(\ii)}$ are determined by Eqs.\,(\ref{eq:d1_pm_d2}) with $\oo = \oo_\D$ for given slab parameters, and in case (ii) the spacings $\ell^{(1)},\ell^{(2)}$ in the supertrimer are determined by the same equations (replacing $d_{1,2} \to \ell^{(1),(2)}$) with $\oo = \oo_\D'$ for the computed superunit characteristics.
Note that, with the method based on combining emitter or absorber units, $\V_1^{(1)}$ ($\V_2^{(1)}$) is made emitter (absorber) at \textit{both} $\oo_\D$ and $\oo_\D'$, while $\V_1^{(2)}~(\V_2^{(2)})$ is made emitter (absorber) at $\oo_{\D}$ and absorber (emitter) at $\oo_{\D'}$.
The desired action of each unit at two frequencies is enabled by further splitting the $\V_j^{(\ii)}$ into equally thick left and right parts of different materials (see inset of Fig.~\ref{fig2}\,(a)), which sufficiently enlarges the parameter space. 

In Figs.~\ref{fig2}\,(b) and (c), the vanishing of $|\aa| + |\dd|$ signifies the occurrence of self-dual operation as $\oo$ crosses the real axis at $\oo_\D$ and $\oo_\D'$, respectively, in dependence of the distance $\ell^{(1)}$ between the superunits $\R^{(1)}$ and $\R^{(2)}$ of a chosen configuration.
Clearly, the self-duality of the total setup at $\oo_\D$ is unaffected by $\ell^{(1)}$ (or $\ell^{(2)}$, not shown), since each superunit is separately self-dual.
In contrast, self-duality occurs recurrently in varying $\ell^{(1)}$ whenever the supertrimer version of Eqs.\,(\ref{eq:d1_pm_d2}) is fulfilled, in similarity to the simple trimer in Fig.~\ref{fig1}\,(e).
As an example in experimentally relevant scaling, the self-dual operation is set at frequencies $\nu_\D^{(\prime)} = \omega_\D^{(\prime)}/2\pi = 600 \, (750)$ THz for a length unit $a = 12.73 \,\mu\text{m}$. 
The same scheme can be implemented to obtain more than two prescribed frequencies with self-dual operation, at the cost of a larger parameter space.

Without the concept of combining emitters and absorbers, the presented application would demand the direct simultaneous solution of four complex nonlinear equations $\aa(\oo_D) = 0,~\aa(\oo_D')=0,~\dd(\oo_D) = 0,~\dd(\oo_D') = 0$ for the total TM.
With the numerical complexity to find such solutions typically increasing exponentially in parameter space dimensionality \cite{Press1992,Boehm2001}, determining realistic parameters for the whole setup at once is enormously harder than for the individual units. 
The procedure described above allows for the recursive composition of the total structure from such simple units which are easily designed individually to yield the required global properties. 

At this point it is essential to comment on the robustness of the self-dual action under the influence of possible realistic perturbations. As indicated in Ref.~\cite{Ge2017}, even small deviations from the ideal parameter values where self-dual action is observed, prevent the divergence of the $S$-matrix eigenvalue. Even though experimentally such deviations are unavoidable, signatures of the effect can be observed, i.e. in the output intensity contrast between the absorbing and emitting states. Such remaining signatures are also present in the case of a small disorder in the refractive indices and the slab lengths, which can simulate the effect of fabrication errors~\cite{Ge2017}. The deviations between the theoretical and the experimental results is also discussed in a very recent experimental realization of CPA-laser action in single cavity~\cite{Wong2016}. Nevertheless, a strong signature of the effect is also observed there. Another factor which could affect the observation of the self-dual action is the incoming wave's deviation from normal incidence by a small angle. In this case, tilted wave incidence can be effectively mapped to rescaled refractive indices in the setup~\cite{Sarisaman2017}. However, a slightly tilted angle of incidence does not suppress the self-dual action significantly.

{\it Directional transparency control.}---The assembly of self-dual setups at selected frequencies from emitters and absorbers above relies on composing partial TM with diagonal zeros into a total one with zero diagonal $\aa,\dd$.
A different composition of such matrices, however, may produce a total TM with one or both \textit{antidiagonal} elements $\bb,\gg$ being zero at the given frequency.
With the reflection amplitudes of the system given by the diagonal elements $s_{11} = \gg/\aa, s_{22} = - \bb/\aa$, this corresponds to unidirectional or bidirectional transparency, respectively.
Those properties are readily provided at preselected frequencies by the proposed design procedure for self-dual setups. 
Consider, e.\,g., the removal of $\R^{(3)}$ in the setup of Fig.~\ref{fig2}\,(a).
Then at $\oo_\D$ the remaining setup $\R^{(1)}\R^{(2)} \sim \D[\omega_\D]\D[\omega_\D]$ has a diagonal total TM
$\footnotesize \arraycolsep=1pt\def\arraystretch{.6}  
\begin{pmatrix} 0 & \bb^{(1)}  \\ \gg^{(1)} & 0 \end{pmatrix} 
\begin{pmatrix} 0 & \bb^{(2)}  \\ \gg^{(2)} & 0 \end{pmatrix} = 
\begin{pmatrix} \aa & 0  \\ 0 & \delta  \end{pmatrix}$,
corresponding to bidirectional transparency. 
This is manifest in Fig.~\ref{fig2}\,(d) where both reflection coefficients $|s_{11}|^2$ and $|s_{22}|^2$ vanish at $\oo_\D$.
In contrast, at $\oo_\D'$ we have $\R^{(1)}\R^{(2)} \sim \E\A[\oo_\D']$ with a TM
$\footnotesize \arraycolsep=1pt\def\arraystretch{.6}  
\begin{pmatrix} 0 & \bb^{(1)}  \\ \gg^{(1)} & \dd^{(1)} \end{pmatrix} 
\begin{pmatrix} \aa^{(2)} & \bb^{(2)}  \\ \gg^{(2)} & 0 \end{pmatrix} =  
\begin{pmatrix} \aa' & 0  \\ \gg' & \dd' \end{pmatrix}$,
corresponding to unidirectional transparency for incidence from the right only, $|s_{11}|^2 = 0 \neq |s_{22}|^2$, as seen in Fig.~\ref{fig2}\,(e).
Clearly, the reversed configuration $\R^{(2)}\R^{(1)}$ would be transparent from the left.
The above simple example reveals a link between singular scattering and directional transparency at a given frequency:
A dimer of an emitter and an absorber is always unidirectionally transparent, while a dimer of two emitter-absorber units is always bidirectionally transparent. As an interesting analogy we mention  the design of lasing and absorbing setups by pairs of unidirectionally invisible parts  in Ref.~\cite{Mostafazadeh2014PRA}.

{\it Conclusions.}---We developed a methodology for the construction of non-Hermitian scattering systems with desired emitting, absorbing, and transparency properties. 
In particular, we demonstrated the systematic design of `absorber', `emitter', and `self-dual' emitter-absorber operation (corresponding to CPA, laser, and CPA-laser in optics) as well as bidirectional and unidirectional transparency at multiple selected frequencies by the recursive assembly of emitter and absorber units.
Based on the structure of the subunit TM, the proposed procedure circumvents the computationally demanding task to search for realizable solutions in high-dimensional parametric spaces, given the experimental limitations for corresponding parameter values.
While focusing on optical scattering in dielectric multilayer structures, the method is applicable to generic wave propagation in systems consisting of localized scattering units.
Due to its versatility and implementation simplicity, based on combinations of single emitter and absorber units, the proposed approach may find possible application in, e.\,g., optical circuitry, and contribute to the advancement of system design techniques.

{\it Acknowledgments.}---P.A.K acknowledges financial support from IKY Fellowships of Excellence for Postdoctoral Research in Greece - Siemens Program.

\end{document}